\documentclass[letterpaper,12pt]{amsart}
	
	\usepackage{amsmath, amsfonts, amsthm, mathrsfs, amssymb, graphicx, subfigure}

	\usepackage{amsaddr}
	
	\theoremstyle{plain}

	\title{Statistical Mechanics of Coulomb Systems: From Electrons and Nuclei to Atoms and Molecules}
	\author{Joel L. Lebowitz}
	\address{Departments of Mathematics and Physics\\
	Rutgers University}

\begin{document}


	
	
	\maketitle
	
	

	
	\begin{abstract}
	This article is dedicated to Elliott Lieb in celebration of his 90th birthday. I recount briefly some history of our joint work on the existence of the thermodynamic limit for Coulomb systems and discuss, even more briefly, two open problems of fundamental nature.
	\end{abstract}
	
	\section{Introduction}
	
	My friendship with Elliott goes back more than six decades. During that period we have collaborated on many scientific projects, including, most significantly, the proof of the existence of the thermodynamic limit.
	
	The first paper on Coulomb systems by Elliott and me was submitted for publication on February 4, 1969 and appeared in the March 31, 1969 issue of Physical Review Letters. I do not remember any referee reports or if we had to make any changes but I do remember (somewhat vaguely) sending a preprint to Freeman Dyson which included some remarks about possible extensions to spin-spin interactions. Dyson responded by noting that there is no lower bound (see \eqref{eq1} below) on these interactions. We did not mention these in our paper. In any case our paper was published pretty fast \cite{LL69}.
	
	The first sentence of that paper reads: ``In this note we wish to report the solution to a classic problem lying at the foundation of statistical mechanics.'' The statement of the result is given in the abstract. ``It is shown that a system made up of nuclei and electrons, the constituents of ordinary matter, has a well-defined statistical-mechanically computed free energy per unit volume in the thermodynamic bulk limit. This proves that statistical mechanics, as developed by Gibbs, really leads to a proper thermodynamics for macroscopic systems.''
	
	The final paragraph of that paper states: ``Many of the ideas presented here had their genesis at the Symposium on Exact Results in Statistical Mechanics at Irvine, California, in 1968, and we should like to thank our colleagues for their encouragement and stimulation: M. E. Fisher, R. Griffiths, O. Lanford, M. Mayer, D. Ruelle, and especially A. Lenard.'' The conference in Irvine in the fall of 1968 was a memorable one. It occurred in the midst of a tumultuous year marked by student uprisings on many campuses around the world. Passing through Paris during that Summer I remember the posters at Orsay demanding ``No Exams''. Irvine was peaceful however and Elliott and I enjoyed our discussions and the sandy beach.
	
	The complete proof of the result was published more than three years later in Advances in Mathematics \cite{LL72}. That paper, which runs to more than 70 pages, also contains an Appendix by Barry Simon on ``Operator Theory Needed in Statistical Mechanics''.
	
	To give some background to our 1968 work let me quote from a review, Statistical Mechanics: A Review of Selected Rigorous Results, which I wrote in December, 1967. It appeared in the Annual Reviews of Physical Chemistry in 1968 \cite{L68}. I review there the proofs of the existence of the thermodynamic limit available at that time. I then note that these proofs require the system Hamiltonian to satisfy the following two conditions:
	\begin{quote}
	(1) H-stability: The Hamiltonian of the system consisting of \(N\) particles has to have a lower bound on its ground state energy proportional to \(N\), i.e.
	\begin{align}\label{eq1}
	H\geq-BN,\quad B\geq0.
	\end{align}
	\end{quote}
	A lower bound on the potential energy of the form \eqref{eq1} implies H-stability but not the converse.
	\begin{quote}
	(2) Temperance: The interaction energy between a group of \(N_1\) particles and another group of \(N_2\) particles separated by a distance \(r\) must fall off with distance faster than \(1/r^d\), where \(d\) is the space dimension, \(d=3\) here (unless otherwise specified).
	\end{quote}
	I then go on to give an outline of the proof by Ruelle and Fisher for systems satisfying conditions 1 and 2. The concluding paragraph of this section reads, with some modifications: ``the only --- and this is perhaps the most important --- system for which the existence of the thermodynamic limit has not yet been proven is a system of charged particles interacting through Coulomb forces only. We certainly expect thermodynamics to apply when the system is electrically neutral, since actual matter consists just of such charged point particles. Nuclear forces are presumably not essential for thermodynamics to exist, and they would give the wrong order of magnitude for binding energies of matter. The existence of a lower bound [eq \eqref{eq1}] for charged particles with hard cores was proven long ago by Onsager. This was recently generalized somewhat by Fisher \& Ruelle. A beautiful proof of condition \eqref{eq1} for a quantum system of charged point particles, at least one species of which obeys Fermi-Dirac statistics, was recently given by Dyson and Lenard. The Fermi-Dirac statistics, i.e. the exclusion principle is essential here, otherwise \eqref{eq1} is violated and \(E_0\sim- N^{7/5}\) and there is no thermodynamic limit.'' \footnote{E. Wigner raised the question at a seminar I gave on this subject: why is not there a spontaneous creation of an over-all neutral \(N\) charged mesons (Bosons) which would have a net negative energy?}
	
	I will not try to describe here how we overcame the problem of long range interactions. The fact that the Coulomb potential decays exactly as \(1/r\) in three dimensions is crucial in our analysis. This permitted us to use Newton's Theorem, derived for the gravitational potential. Applied to Coulomb systems with positive and negative charges it states that a spherically symmetric charge neutral system contained in a sphere of radius \(R\) produces no electric field outside \(R\). We really enjoyed giving a reference to Newton in our paper.
	
	At this point I would like to mention how Onsager proved H-stability for a classical system of charged hard spheres of diameter \(D\) as this may not be well known. Onsager's proof also involves the use of Newton's Theorem. It starts by noting that in the presence of hard cores, which requires that the distance between two particles \(|x_i-x_j|\) must be greater than \(D\), ``the charge on each particle can be considered (as far as the interactions are concerned) to be smeared out on the surface of a sphere of diameter \(D\). This has a finite self energy \(\varepsilon(D)\). It is then a basic fact of electrostatics that
	\begin{align}
	&\frac{1}{2}\sum_{i\ne j}e_ie_j/|x_1-x_j|=\frac{1}{2}\int_{\mathbb R^3}E^2(x)\mathrm d^3x-\sum\text{ self energy}\\
	&\qquad\ge-N\varepsilon(D),\quad |x_1-x_j|\ge D.\nonumber
	\end{align}
	Here \(\mathbf E(x)\) is the electric field at the point \(x\) and the integral is obviously non-negative.''
	
	The above is taken from an article I wrote in 1980 \cite{L81}. The article is based on lectures I gave at the International School of Mathematical Physics, ``Ettore Majorana'', Erice, Sicily, June, 1980. This school was organized by Giorgio Velo and Arthur Wightman and was attended by Elliott, J\"urg Fr\"ohlich and many other colleagues. It was a most enjoyable school with visits to historical Greek and Roman sites.
	
	I will devote the rest of this article to two of the questions Elliott and I discussed at that meeting which are, as far as I know, still open and interesting.

	\section{Thermodynamics limit for dipoles}
	
	One of the most interesting still open problems which Elliott and I discussed a lot during the meeting at Erice and in other places is the existence of the thermodynamic limit for systems of dipoles, \nobreak{located} on some regular lattice, like \(\mathbb Z^3\), or having hard cores in \(\mathbb R^3\). The ``bare'' dipole interaction potential of particles with a permanent dipole moment of strength \(|\mu|\) in an external field \(\mathbf E\) is given by
	\begin{align}
	U_\alpha=\frac{1}{2}\sum_{i\ne j}[\mu_i\cdot\mu_j-3(\mu_i\cdot\mathbf{\hat r}_{ij})(\mu_j\cdot\mathbf{\hat r}_{ij})]/r^3_{ij}-\sum\mathbf E\cdot\mu_i,
	\end{align}
	where \(\mu_i\) are the dipole moments, vectors in \(\mathbb R^3\), \(\mathbf r_i\in\Lambda\subset\mathbb Z^d(\mathbb R^d)\) is the position of the \(i\)-th particle and \(\mathbf{\hat r}_{ij}\) is the unit vector along \((\mathbf r_i-\mathbf r_j)\). There can also be other short range potentials such as hard cores for \(\mathbf r_i\in\mathbb R^3\), which assures H-stability. The dipole-dipole interaction falls off as \(r^{-3}\), just failing to satisfy the temperance condition in three dimensions, \(\mathbf E\) is an external electric field.
	
	It was shown by Griffiths and by Fr\"ohlich and Park \cite{FP78}, and later by Banerjee, Griffiths and Widom in \cite{BGW98}, that when \(\mathbf E=0\) this system has a well defined thermodynamic limit as \(\Lambda\to\mathbb Z^d(\mathbb R^d)\) which is shape independent, both classically and quantum mechanically. Since \(\mathbf E=0\) the average net polarization is equal to zero.
	
	When \(\mathbf E\ne0\) on the other hand one expects a net polarization and a shape dependent free energy density which can be computed from macroscopic considerations for elliptically shaped domains \(\Lambda\). Proving this remains a challenge to mathematical physics. I have not found any results on this problem later than 1998, the date of \cite{BGW98}.
	
	When the dipoles are confined to a lower dimensional region the \(r^{-3}\) decay is sufficiently rapid to permit the standard proofs of the existence of the thermodynamic limit to go through. The actual properties of such systems in one and two dimensions is a rich topic of investigation. The structure of the ground state of such a dipole system in two dimensions with a limited number of orientations was discussed in a work by Giuliani, Lebowitz and Lieb \cite{GLL07}. There ``We prove that a system of discrete two-dimensional (2D) in-plane dipoles with four possible orientations, interacting via a three-dimensional (3D) dipole-dipole interaction plus a nearest neighbor ferromagnetic term, has periodic striped ground states. As the strength of the ferromagnetic term is increased, the size of the stripes in the ground state increases, becoming infinite, i.e., giving a ferromagnetic ground state, when the ferromagnetic interaction exceeds a certain critical value. We also give a rigorous proof of the reorientation transition in the ground state of a 2D system of discrete dipoles with six possible orientations, interacting via a 3D dipole-dipole interaction plus a nearest neighbor antiferromagnetic term. As the strength of the antiferromagnetic term is increased, the ground state flips from being striped and in plane to being staggered and out of plane. An example of a rotator model with a sinusoidal ground state is also discussed.'' This paper is part of a series of works by Alessandro, Elliott and me on periodic structures. For additional work in this direction see Fr\"ohlich and Spencer \cite{FS81} and Giuliani \cite{G09}.

	\section{From electrons and nuclei to atoms and molecules}
	
	Having established the existence of the thermodynamic limit there is little doubt that an appropriate fundamental description of bulk macroscopic matter in equilibrium is via the Gibbs density matrix \(\rho\sim\exp(-\beta H)\), with \(H\) the Coulomb Hamiltonian of nuclei and electrons. It is then natural to ask for the actual structure of a macroscopic system consisting of electrons and nuclei. We know from empirical observations that this bare Hamiltonian, with the right statistics, can lead to the formation of many different states of matter. The simplest of these are gases and liquids. In these cases the bare Coulomb interaction is replaced by an effective Hamiltonian \(H_{\mathrm{eff}}\) whose basic entities are atoms or molecules.
	
	For example, to obtain the properties of helium or nitrogen at moderate temperatures and pressures it certainly suffices, for all practical purposes to consider the atoms or molecules as the basic units with an effective two- or three-body interaction between them. There will always be a certain fraction of ``ionized'' atoms corresponding to ``free'' electrons and ions. This ``degree of ionization'' will in fact be complete when the density goes to zero at a fixed positive temperature [7]. The mathematical-physics issue is then to obtain an \(H_{\mathrm{eff}}\) which empirically seems to be approximately the same for a large range of temperatures and densities. The degree of ionization increases with temperature and the system becomes a plasma at sufficiently high temperature. It is also true that as the density increases the degree of ionization also increases leading to a phase transition from an insulator to a conductor.
	
	A proof of the formation of atoms and molecules, starting with the Coulomb Hamiltonian, was achieved so far only for special limits of zero pressure and zero temperature, by C. Fefferman \cite{F85} and by J. Conlon, E. Lieb and H. T. Yau \cite{CLY89}, where the approximate formalism has been given a rigorous foundation. The result by Fefferman (established modulo some reasonable assumptions) in the simplest case of the electron-proton system is as follows: if one fixes the chemical potential \(\mu=\frac{1}{2}(\mu_\mathrm{e}+\mu_\mathrm{p})\) below the ground-state energy of a hydrogen atom \(E_\mathrm{at}\), then in the limit \(\beta\to\infty\) the system consists of free electrons and protons. On the other hand, when the chemical potential is fixed slightly above \(E_\mathrm{at}\), the system will consist of independent hydrogen atoms when \(\beta\to\infty\). Conlon, Lieb and Yau extended this result to molecules.
	
	In the above limits both the particle density and the temperature go to zero. One therefore has to go beyond this formalism to gain an understanding of the structure of ordinary fluids at moderate temperature and density in which almost all electrons and nuclei are bound in neutral clusters of atoms or molecules with a certain small degree ionization as described by the Saha equation.
	
	A fundamental conceptual difficulty in dealing with these problems is the lack of an \emph{a priori} distinction between ``free or ionized'' and ``bound or atomic'' states in the many-body quantum formalism. We discussed this problem in a paper with Macris and Martin in 1992 \cite{LMM92}. We write there that to overcome this problem Girardeau \cite{G90} suggested that the bound and free electron states for partially ionized hydrogen are given in terms of the spectrum of the equilibrium one-proton-one-electron density matrix. It associates the discrete part of the spectrum with ``bound states'' and the continuum with ``ionized states''. This was also suggested in an earlier work by Macris and Martin. The most recent review of the subject that I am aware of is \cite{MM94}.
	
	These ideas were studied for some very simple model systems in \cite{LMM92} and \cite{MM94}. They seem worthwhile investigating further as are other approaches to this problem. The extension of the results of Fefferman and of Conlon, Lieb and Yau to finite density and temperature are most desirable.

	\section{Acknowledgements}
	
	I would like to thank Alessandro Giuliani, Robert Seiringer and August Krueger for helpful comments and assistance with the manuscript. My work on Coulomb systems was supported by the AFOSR and the NSF.


	\end{document}